\documentclass[twocolumn,aps,pra]{revtex4-1}
\usepackage{amsmath,amssymb,amsfonts,bbm,graphicx,color,times}
\usepackage[english]{babel}
\usepackage{amsmath,graphicx,bbm}
\usepackage{geometry}
\usepackage{float}

\catcode`\|=\active \def|{
\fontencoding{T1}\selectfont\symbol{124}\fontencoding{\encodingdefault}}
\newcommand{\bignone}{}
\newcommand{\mathd}{\mathrm{d}}
\newcommand{\mathe}{\mathrm{e}}
\newcommand{\tmem}[1]{{\em #1\/}}
\newcommand{\tmop}[1]{\ensuremath{\operatorname{#1}}}
\newcommand{\tmtextbf}[1]{{\bfseries{#1}}}

\begin{document}

\title{Experimental investigation of the effect of classical noise\\
  on quantum
non-Markovian dynamics}

\author{Simone Cialdi}
\email{simone.cialdi@unimi.it}

\author{Claudia Benedetti}
\author{Dario Tamascelli}

\author{Stefano Olivares}
\author{Matteo G. A. Paris}

\author{Bassano Vacchini}
\email{bassano.vacchini@mi.infn.it}

\affiliation{Quantum Technology Lab, Dipartimento di Fisica ``Aldo Pontremoli'', Universit{\`a} degli
Studi di Milano, Via Celoria 16, I-20133 Milan, Italy}

\affiliation{INFN, Sezione di Milano, Via Celoria 16, I-20133 Milan, Italy}

\begin{abstract}
  We provide an experimental study of the relationship between the action of
  different classical noises on the dephasing dynamics of a two-level system
  and the non-Markovianity of the quantum dynamics. The two-level system is
  encoded in the photonic polarization degrees of freedom and the action of the
  noise is obtained via a spatial light modulator, thus allowing for
  an easy engineering of
  different random environments. The quantum non-Markovianity of the dynamics
  driven by classical Markovian and non-Markovian noise, both Gaussian and
  non-Gaussian, is studied by means of the trace distance. Our study clearly
  shows the different nature of the notion of non-Markovian classical process
  and non-Markovian quantum dynamics.
\end{abstract}

\maketitle

\section{Introduction}
The characterization of quantum non-Markovian processes has recently
attracted a lot of attention: besides its conceptual interest it
might open the way to obtain improved performances in quantum
thermodynamics \cite{Binder2018}, higher sensitivities in quantum
metrology \cite{Chin2012a,Haase2018b} and techniques for complex quantum
systems probing \cite{Smirne2013a}.  A natural and intriguing question
is the relationship between the proposed notions of quantum
non-Markovian dynamics (see \cite{Rivas2014a,Breuer2016a,Devega2017a}
for reviews) and the classical notion of memoryless or Markovian
process. This parallel has been the object of different theoretical
studies \cite{Vacchini2011a,Vacchini2012a,Guarnieri2014a,Gessner2014b,Li2018a,Smirne2019a} and
has indeed provided the motivation for one of the first approaches to
the problem \cite{Lindblad1979a}.

In this paper we address this question from a new viewpoint, relying
on the experimental realization of quantum dynamics depending on a
classical random processes. In such a way we relate a classical input
with a quantum output and investigate the features of the latter with
respect to the former. To this aim we need to generate a wide variety
of classical processes with known features. We perform this task by
obtaining such stochastic processes as solution of suitable stochastic
differential equations, so that we can obtain both Gaussian and
non-Gaussian, Markovian and non-Markovian classical processes.

The experimental implementation is based on a quantum optics setup
which allows to engineer in a controlled way a dephasing dynamics
determined by a classical stochastic process, which affects the
polarization degrees of freedom of photons. A suitable configuration
allows to address in parallel a high number of realizations of the
process, and average them automatically in the detection stage.

The paper is organized as follows. In Sect.~\ref{modello} we introduce
the model of the system and the environment used to relate a classical input
with a quantum output. In Sect.~\ref{noise} we discuss how to
generate classical noises with known features. We introduce the
experimental setup in Sect.~\ref{experiment-app} and discuss the
obtained results in Sect.~\ref{exp-res}. We draw our conclusions and
final remarks in Sect.~\ref{ceo}.

\section{The model}
\label{modello}

\subsection{Evolution map}

In order to investigate by means of experiment how the different hallmarks of
classical noise, such as being Gaussian or Markovian, affect the features of a
quantum dynamics, especially in view of the property of non-Markovianity, we
consider the following simple but versatile model. The Hamiltonian describing
the time evolution of the system is given by
\begin{eqnarray}
  H ( t ) & = & X ( t ) \hbar \omega_{0} \sigma_{z} ,  \label{eq:H}
\end{eqnarray}
where $X ( t )$ denotes a classical stochastic process with independent
increments, describing the effect of the environmental noise on the two-level
system of interest, and $\omega_{0}$ denotes the natural energy splitting of
the two-level system, fixing scale and dimensions. The time evolution operator
determining the effect of each single realization of the noise reads
\begin{eqnarray}
  U ( t,0 ) & = & \mathe^{-i \omega_{0} \sigma_{z} \int^{t}_{0} \mathd \tau X
  ( \tau ) \bignone} , \nonumber
\end{eqnarray}
so that upon averaging over the environmental influence one obtains for the
reduced system dynamics

\begin{eqnarray}
  \rho_{S} ( t ) & = & \Lambda ( t ) [ \rho_{S} ( 0 ) ] \nonumber\\
  & = & \mathbbm{E_{}} [ U ( t,0 ) \rho_{S} ( 0 ) U ( t,0 )^{\dag} ] ,
  \nonumber
\end{eqnarray}
where $\mathbbm{E [ \cdot ]}$ denotes the expectation value over the sample
space of the noise. We are interested in investigating the behavior of
$\rho_{S} ( t )$ in its dependence from the noise $X ( t )$.

To this aim we denote by
\begin{eqnarray}
  \mathcal{X} ( t ) & = & \int^{t}_{0} \mathd \tau X ( \tau )  \label{eq:int}
\end{eqnarray}
the integral over time of the stochastic process, leading to $U ( t,0 ) =
\tmop{diag} ( \mathe^{-i \omega_{0} \mathcal{X} ( t )} , \mathe^{+i \omega_{0}
\mathcal{X} ( t )} )$, so that the reduced system dynamics is fully captured
by the transformation of the off-diagonal matrix element
\begin{eqnarray}
  \rho^{10}_{S} ( 0 ) & \xrightarrow{t} & \rho^{10}_{S} ( 0 ) \mathbbm{E} [
  \exp ( -2i \omega_{0} \mathcal{X} ( t ) ) ] .  \label{eq:phase}
\end{eqnarray}
This simple description allows for a high experimental freedom in the
implementation of the environmental noises and a unique characterization of
non-Markovianity of the ensuing quantum dynamics. Indeed the evolution
corresponds to a pure dephasing dynamics, for which all definitions of quantum
non-Markovianity coincide {\cite{Rivas2014a,Breuer2016a,Devega2017a}}, so that
a clearcut signature of quantum memory effects can be provided. In such a way
the considered model well serves the purpose of exploring the effects of
different classical noises on a quantum dynamics.

\subsection{Features of quantum dynamics}
\label{introTD}

The signature of the quantum dynamics that we want to address, in its
dependence on the noise describing the effect of the environment, is its
non-Markovianity. In the considered model, once we fix the noise acting
in the Hamiltonian, we obtain a quantum dynamics which, while being given by a
unitary transformation for each fixed realization of the noise, provides a
time dependent collection of completely positive trace preserving maps upon
averaging over the noise. This collection of maps describes a stochastic
quantum dynamics, which would generally arise as a consequence of the
interaction with a suitable quantum environment. Indeed, as it has been
recently considered, any classical average can be seen to arise from a
microscopic description with ancillary quantum degrees of freedom initialized
in a classical state {\cite{Breuer2018a}}. In this respect the obtained
dynamics provides a well-defined reduced quantum dynamics, whose features can
be studied in view of the relationship between the properties of the input
noise and the features of the output maps. As a figure of merit we
will consider the non-Markovianity of the ensuing quantum dynamics as
characterized by the behavior in time of the trace distance between two
initially distinct system states. This viewpoint was first introduced in
{\cite{Breuer2009b,Laine2010a}} and connected to a notion of information
exchange between system and environment. In particular, it is important to
stress that this physical intuition remains confirmed even if the map, as in
this case, is obtained upon averaging with respect to a classical
label. This is an
important issue, which has been the object of different investigations
{\cite{Chruscinski2013a,Megier2017a,Breuer2018a}}. For the present simple
model, as already stressed, essentially all proposed definitions of
non-Markovianity coincide {\cite{Breuer2016a}} since the dynamics is captured
by the transformation Eq.~(\ref{eq:phase}). The natural quantifier of
non-Markovianity for this class of models is therefore the behavior of the
quantity
\begin{eqnarray}
  D [ \{ X ( t ) \} ] & = & | \mathbbm{E} [ \exp ( -2i \omega_{0} \mathcal{X}
  ( t ) ) ] | ,  \label{eq:decoh}
\end{eqnarray}
which describes the dephasing effect of the environmental noise on the system
dynamics. In the trace distance formalism this estimator is obtained considering
a pair of states which maximize the possible revivals of the considered
quantity. A monotonic decrease in time of this quantity corresponds to a
Markovian dynamics, while a non-Markovian dynamics is obtained if revivals in
time appear.

\section{Classical noise generation\label{noise}}

In considering classical stochastic processes, two classes stand out in their
relevance for applications and theoretical treatments, namely Gaussian and
Markovian processes. In both cases a relatively simple description applies, at
variance with the case of a generic process. As a matter of fact, while in the general case
a description of the process calls for knowledge of all its correlation
functions, fixing the probability for given outcomes of the random variable at
given times, in the case of Gaussian and Markovian processes a drastic
simplification applies. A Gaussian process is in fact fixed by
first and second moments only, while a Markovian process is determined by
initial probability distribution and transition probability
{\cite{Gardiner2004a}}. It is therefore of interest to explore the effect of
noise on a quantum dynamics classifying the classical noises with respect to
these two distinctive features.

To the aim of generating in a simple way these different type of noises we
consider as starting point two Markovian processes whose realizations can
be easily simulated. Our starting tools are therefore Wiener 
and random
telegraph noise, both Markovian: Gaussian the former, non-Gaussian the latter.
Stochastic processes $X ( t )$ with different features to be used in the
dynamics given by Eq.~(\ref{eq:H}) will be obtained as solution of stochastic
differential equations with different input noises.

As Markovian Gaussian process we will consider a Ornstein-Uhlenbeck process
$X_{\tmop{OU}} ( t )$ with friction coefficient $\gamma$ and diffusion
constant $\sigma$, which can be obtained as solution of a linear stochastic
differential equation driven by Wiener noise. We will further denote as
$X_{\tmop{RTN}} ( t )$ a random telegraph noise with switching rate $\gamma$
and step one, whose realizations therefore jump from plus to minus one
according to a Poisson process with rate $\gamma$. Such a process is still
Markovian, but its probability density is not in Gaussian form. Using the
realization of these processes as input noise we can obtain noises with
different features.

Let us first consider the equation
\begin{eqnarray}
  dY ( t ) & = & - \kappa Y ( t ) \tmop{dt} +dX_{\tmop{OU}} ( t ) , 
  \label{eq:ii}
\end{eqnarray}
with $\kappa$ a positive rate. The process $Y ( t )$ is still Gaussian due
to linearity of the equation but non-Markovian, since its determination
requires the knowledge of $X_{\tmop{OU}} ( t )$ up to time $t$
{\cite{Fox1977a,Machado1984a}}. Using the same strategy we can obtain the
increments of a process which is neither Gaussian nor Markovian considering
the stochastic differential equation
\begin{eqnarray}
  dZ ( t ) & = & - \mu Z ( t ) \tmop{dt} +dX_{\tmop{RTN}} ( t ) , 
  \label{eq:iii}
\end{eqnarray}
with $\mu$ a positive rate. Using as seeds Wiener and random
telegraph noise we are thus able to generate, via the stochastic differential
equations given by Eq.~(\ref{eq:ii}) and Eq.~(\ref{eq:iii}), increments of
stochastic processes which share or lack the distinct features of Gaussianity
and Markovianity according to the corresponding well established classical
definitions.

In order to estimate the effect of the different noises on the dynamics we
further need to evaluate the expectation value of the integral over time of
the considered noise, defined as in Eq.~(\ref{eq:phase}). The analytic
evaluation of this quantity is only feasible in special cases. For the case of
the Ornstein-Uhlenbeck process we define
\begin{eqnarray}
  \mathcal{X}_{\tmop{OU}} ( t )_{} & \equiv & \int^{t}_{0} \mathd \tau
  X_{\tmop{OU}} ( \tau ) , \label{fantasy1}
\end{eqnarray}
and exploiting Gaussianity one obtains for the quantity determining the
dephasing of the two-level system 
\begin{multline}
\!\! \!\!  D [ \{ X_{\tmop{OU}} ( t ) \} ]  \equiv  | \mathbbm{E} [ \exp ( -2i
  \omega_{0} \mathcal{X}_{\tmop{OU}} ( t ) ) ] |  \label{eq:dou}\\
  = \exp \left( - \frac{\omega_{0}^{2} \sigma^{2}}{\gamma^{3}} ( 2 \gamma
  t-3-e^{-2 \gamma t} +4e^{- \gamma t} ) \right) , 
\end{multline}
where $\gamma$ and $\sigma$ denote respectively friction and diffusion
coefficient of the process.

For the case of random telegraph noise, defining on the same footing
\begin{eqnarray}
  \mathcal{X}_{\tmop{RTN}} ( t )_{} & \equiv & \int^{t}_{0} \mathd \tau
  X_{\tmop{RTN}} ( \tau ) , \label{fantasy2}
\end{eqnarray}
one can show that the dephasing factor takes the form
{\cite{Abel2008a}}
\begin{multline}
\!\! \!\!  D [ \{ X_{\tmop{RTN}} ( t ) \} ]  \equiv  | \mathbbm{E} [ \exp ( -2i
  \omega_{0} \mathcal{X}_{\tmop{RTN}} ( t ) ) ] |  \label{eq: drtn}\\
  = \mathe^{- \gamma t} \left[ \cosh (\nu t)
  +
  \frac{\gamma}{\nu} \sinh (\nu t) \right] ,
\end{multline}
with $\nu=\sqrt{\gamma^{2} -4 \omega_{0}^{2}}$.
These explicit expressions allow to estimate the dephasing factor of
Eq.~(\ref{eq:decoh}) and study its monotonicity properties as a function of
time. As we discuss in Sect.~\ref{exp-res} these estimates are indeed
confirmed by the experimental results, and validate the theoretical analysis.
It appears in particular that while the dephasing due to Ornstein-Uhlenbeck
noise $D [ \{ X_{\tmop{OU}} ( t ) \} ]$ is a decreasing function of time for
any value of $\gamma$ and $\sigma$, the contribution corresponding to the random
telegraph noise $D [ \{ X_{\tmop{RTN}} ( t ) \} ]$ can also exhibit an
oscillating behavior {\cite{Benedetti2014a}}. Note that both processes are
examples of Markovian colored noise and have an exponentially decaying
correlation function, namely {\cite{Gardiner2004a}}
\begin{eqnarray}
  \mathbbm{E} [ X_{\tmop{OU}} ( t ) X_{\tmop{OU}} ( s ) ] & = &
  \frac{\sigma^{2}}{2 \gamma} \exp ( - \gamma | t-s | ) \nonumber
\end{eqnarray}
and
\begin{eqnarray}
  \mathbbm{E} [ X_{\tmop{RTN}} ( t ) X_{\tmop{RTN}} ( s ) ] & = & \exp ( -2
  \gamma | t-s | ). \nonumber
\end{eqnarray}
In order to consider a classical non-Markovian process, still retaining the
property of Gaussianity, we refer to Eq.~(\ref{eq:ii}). The relevant quantity
is again $D [ \{ Y ( t ) \} ]$, which can be evaluated exploiting the fact
that $Y ( t )$ is again Gaussian and relying on the properties of the
Ornstein-Uhlenbeck process. The result reads 
\begin{widetext}
\begin{eqnarray}
  D [ \{ Y ( t ) \} ] & = & | \mathbbm{E} [ \exp ( -2i \omega_{0}  \mathcal{Y}
  ( t ) ) ] |  \label{eq:dy}\\
  & = & \exp \left\{ - \omega_{0}^{2} \sigma^{2} \frac{( \gamma - \kappa )^{2} -
  ( \gamma e^{- \kappa t} - \kappa e^{- \gamma t} )^{2} + \gamma \kappa ( 2e^{-( \gamma
  + \kappa )t} -e^{-2 \gamma t} -e^{-2 \kappa t} )}{\gamma \kappa^{} ( \gamma - \kappa
  )^{2} ( \gamma + \kappa )} \right\} , \nonumber
\end{eqnarray}
\end{widetext}
where according to Eq.~\eqref{fantasy1} and \eqref{fantasy2} we have
denoted the integrated process as $\mathcal{Y} ( t )\equiv \int^{t}_{0} \mathd \tau
  Y ( \tau )$.  The dephasing factor shows a monotonic decaying behavior for all
possible values of the constants $\gamma$ and $\kappa$, friction
coefficient of the Ornstein-Uhlenbeck and rate appearing in the
stochastic differential equation Eq.~\eqref{eq:ii} respectively.

The last process that we will consider is the solution of Eq.~(\ref{eq:iii}),
which is neither Gaussian nor Markovian due to the fact that the driving noise
is colored and non-Gaussian. The evaluation of the corresponding dephasing
factor
\begin{eqnarray}
  D [ \{ Z ( t ) \} ] & = & | \mathbbm{E} [ \exp ( -2i \omega_{0}  \mathcal{Z}
  ( t ) ) ] |,  \label{eq:dz}
\end{eqnarray}
with $\mathcal{Z} ( t )\equiv \int^{t}_{0} \mathd \tau
  Z ( \tau )$,
calls for a numerical evaluation since we can no more exploit the important
simplification in the evaluation of the characteristic function warranted for
Gaussian processes. In particular, as confirmed by the experiment, it appears
that depending on the parameter values also in this case an oscillating
behavior can show up.
 It thus appears that in this context non-Markovianity of
the quantum dynamics appears when the relevant classical process is non-Gaussian, rather then being related to a lack of the Markov property.

\begin{figure}
  \includegraphics[width=0.99\columnwidth]{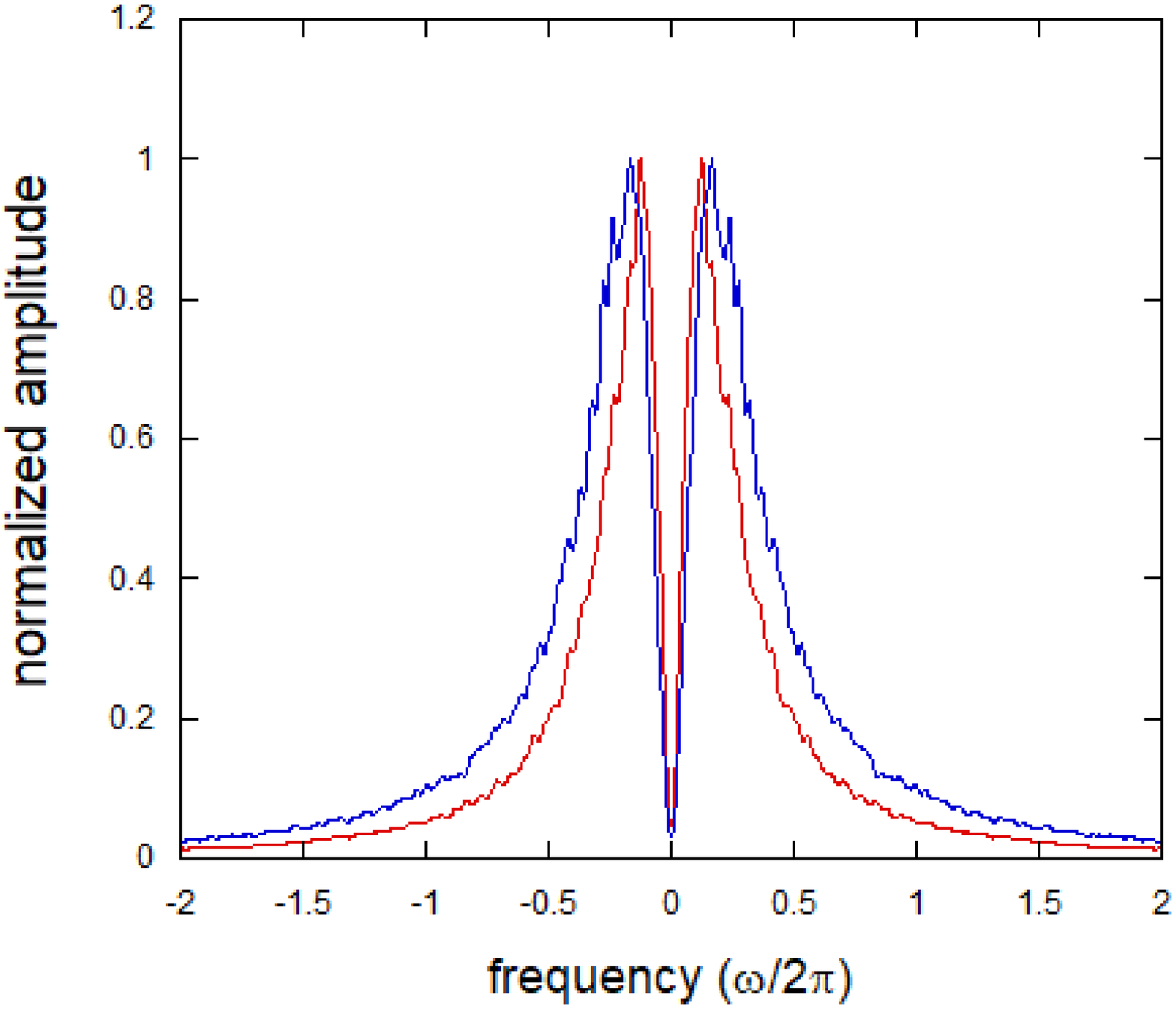}
\caption{\label{fig:spettroZ} Spectrum of the process $Z(t)$ obtained
  by the numerical evaluation of the stochastic
  differential equation Eq.~(\ref{eq:iii}) for the case $\mu=0.5$ (red line) and
  $\mu=1$ (blu line). The  coefficient $\gamma$ characterizing the
  RTN is set to 0.5.
}
\end{figure}

In various theoretical papers and experimental implementations
{\cite{Liu2011a}} it has been shown that the appearance of quantum
non-Markovianity in situations in which the environmental interaction can be
characterized by a spectral density is typically related to a non-trivial peak
structure of the relevant frequency spectrum. In this respect it is natural to
investigate also in the present framework the relationship between spectral
properties of the noise and features of the quantum dynamics. For both
Ornstein-Uhlenbeck and random telegraph noise the spectrum has a Lorentzian
shape centered in zero, corresponding to the exponential decay of the
two-time correlation function. The correlation function of the process $Y ( t
)$ takes instead the form
\begin{widetext}
\begin{multline}
  \mathbbm{E} [ Y ( t ) Y ( s ) ] = \left( \frac{\sigma}{\gamma - \nu}
  \right)^{2} \left\{
    \frac{\gamma}{2} ( \mathe^{- \gamma | t-s |} - \mathe^{-
  \gamma ( t+s )} ) + \frac{\nu}{2} ( \mathe^{- \nu | t-s |} - \mathe^{- \nu (
  t+s )} ) \right. \nonumber\\
+ \left.
  \frac{\gamma \nu}{\gamma + \nu} ( \mathe^{- \nu t- \gamma s}
  + \mathe^{- \gamma t- \nu s} - \mathe^{- \nu | t-s |} - \mathe^{- \gamma |
  t-s |} ) \right\} \nonumber
\end{multline}  
\end{widetext}
The process is only asymptotic stationary, with associated power spectrum
\begin{eqnarray}
  S ( \omega ) & = & \frac{\sigma^{2}}{2 \pi} \frac{\omega^{2}}{( \gamma^{2} +
  \omega^{2} ) ( \kappa^{2} + \omega^{2} )} \label{power}
\end{eqnarray}
featuring a double-peaked structure and a dip at small frequencies.
The same feature is shared by the spectrum of the process $Z(t)$
arising as  solution of Eq.~(\ref{eq:iii}), which can be evaluated
numerically and is shown in Fig.~\ref{fig:spettroZ}.

Despite the non trivial structure of the power spectrum, as follows
from Eq.~(\ref{eq:dy}) the trace distance still exhibits a
monotonically decaying behavior, reflecting a Markovian dynamics, for the $Y(t)$
process, while oscillations may be present for the $Z(t)$ process.  It
therefore appears that in this context the correlation function of the
classical process and the associated power spectrum does not embody
the relevant information in characterizing the memory properties of
the quantum dynamics.

\section{Experimental implementation}
\label{experiment-app}
\begin{figure}
\includegraphics[width=0.99\columnwidth]{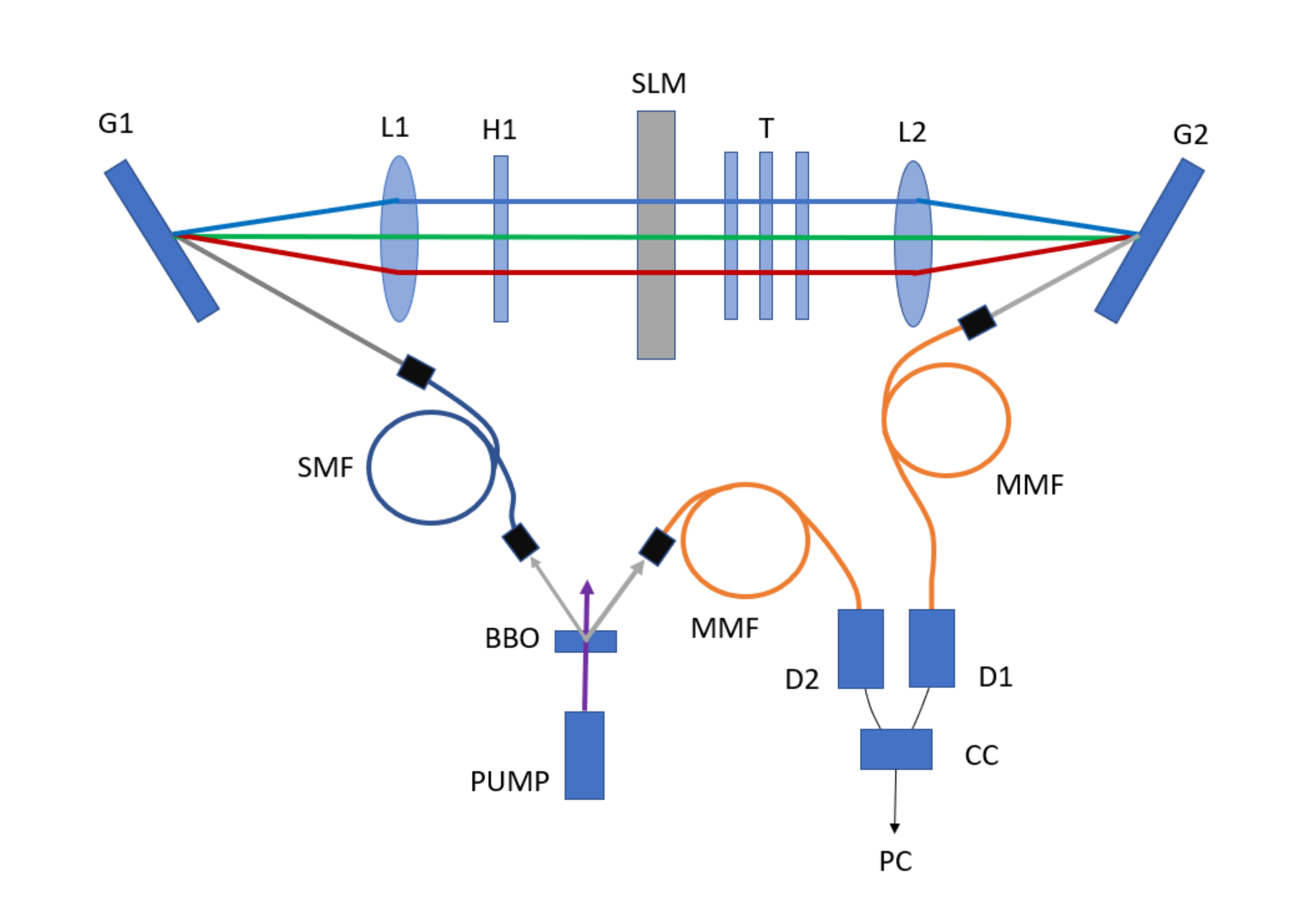}
\caption{\label{fig:apparato} Schematic diagram of our 
apparatus. A couple of frequency-entangled
photons is generated via parametric down-conversion (PDC) through a
BBO crystal,
using a $405.5$nm laser diode as pump.
One photon is sent via a
multi-mode fiber (MMF) to the single-photon detector D2. The other is
sent through a single-spatial-mode and polarization preserving fiber
(SMF) to the 4F system (composed by two diffraction
gratings G1-G2 and two lenses L1-L2). The initial state of the photon is prepared by the half-wave plate H1.
T is the tomographic apparatus, made of a quarter-wave plate, a half-wave
plate and a polarizer.  The photon is then sent through a MMF to the
single-photon detector D1. Finally, an electronic device measures the
coincidence counts (CC) and sends them to the computer (PC).
}
\end{figure}

The effect of a classical noise on a quantum dephasing dynamics can be
experimentally investigated in a quantum optics setup. To this aim we encode the quantum
degrees of freedom in the polarization state of photons and let the
noise affect the phase information. Efficiently generating and averaging over
the different realizations of the noise provides the major obstacle in order
to experimentally study the effect of classical disturbance on a quantum dynamics.
This difficulty can be overcome by exploiting a recently realized all-optical quantum
simulator {\cite{Cialdi2017a}}. This apparatus allows to obtain
many realizations of the considered stochastic process in parallel and directly averages
over them at the detection stage. While details of the experimental setup
have been given in {\cite{Smirne2013a,Cialdi2017a}}, we will here provide the
logical scheme of the apparatus, represented in
Fig.~\ref{fig:apparato}.
The core of the apparatus is a spatial light modulator (SLM) placed in
the Fourier plane between the two lenses L1 and L2 of the 4F
system. The SLM is a 1D liquid crystal mask ($640$ pixels,
$100$ $\mu$ m/pixel) used to introduce a different phase (externally
controlled by the computer) to each pixel, implementing the simulation of the dynamical map. This device
thus imprints different phases depending on the position and on the polarization
state of the incoming photon. In the experimental device photons are
generated by parametric down-conversion and a suitable grating
provides a spatial separation of the different frequency
components. The SLM acts differently on the different spectral
components, thanks to their spatial separation, and thus allows to
encode in parallel different realizations of the noise. This
experimental setup further allows to perform the average over the realizations
of the noise by collecting the different spatial components through
the lens L2 and the grating G2 into a multi-mode fiber (MMF). The detection stage
is in fact performed after recollecting the signal via the MMF, so that one averages over the spectral
components and therefore the different realizations of the noise. We
observe that the parametric down-conversion (PDC) spectrum is selected by the limited width of the
$H1$ plate mount. For this reason we are limited to use $n=100$ out of
the $640$ pixel available on the SLM (which corresponds to $100$
realizations in parallel of the noise).

\begin{figure}[h]
 \includegraphics[width=0.99\columnwidth]{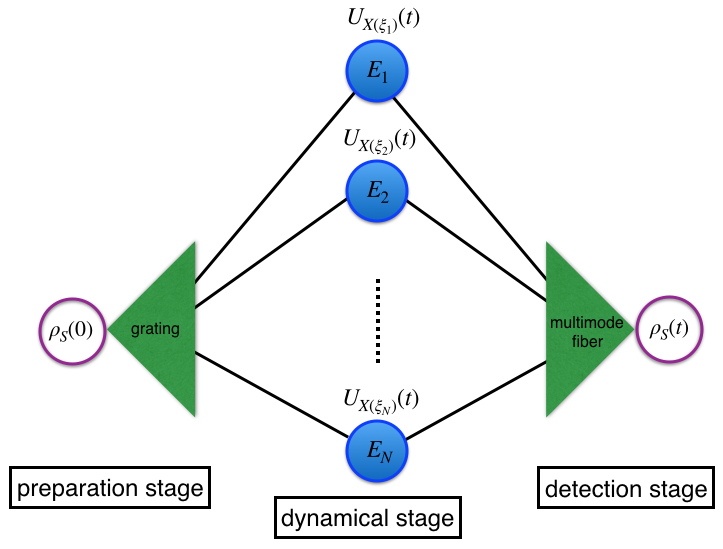}
  \caption{\label{fig:schema}Logical scheme of the experimental setting. The
  preparation stage involves generation of the photons and spatial separation
  of the different spectral components via a grating. The dynamical stage
  involves interaction with different regions of the SLM,
  imprinting different phases depending on the realization of the noise
  associated to the region, corresponding to a Hamiltonian
  interaction $U_{X(\xi)}(t)$ with a fixed noise realization. The detection stage involves recombination of the
  different spectral components by means of a MMF and a final
  photon detection.}
\end{figure}
As shown in the logical scheme Fig.~\ref{fig:schema} this simple experimental
setting nicely reproduces the framework considered in {\cite{Breuer2018a}},
namely the description of the overall reduced dynamics arising as a mixture of
Markovian dynamics. In the present case in particular the system dynamics
which get mixed are given by unitary maps $U_{X(\xi)}(t)$, each characterized by a single
realization of the stochastic process. In the experimental realization of the
scheme it clearly appears how non-Markovianity arises because of the presence
of degrees of freedom dynamically coupled to the observed ones and later
averaged over.

\section{Effect of noise on the quantum dynamics\label{exp-res}}

We here report about the experimental results for the realizations of the
different kind of noises considered in Sec.~\ref{noise}, using the
apparatus described in Sec.~\ref{experiment-app}. In order to generate
the different noises we have numerically solved the stochastic differential
equations considered in Sec.~\ref{noise}. The obtained values have been passed
over to the SLM so as to affect the phase of the photons according to the
dynamics given by Eq.~(\ref{eq:H}). 
\begin{figure}[H]
  \includegraphics[width=0.99\columnwidth]{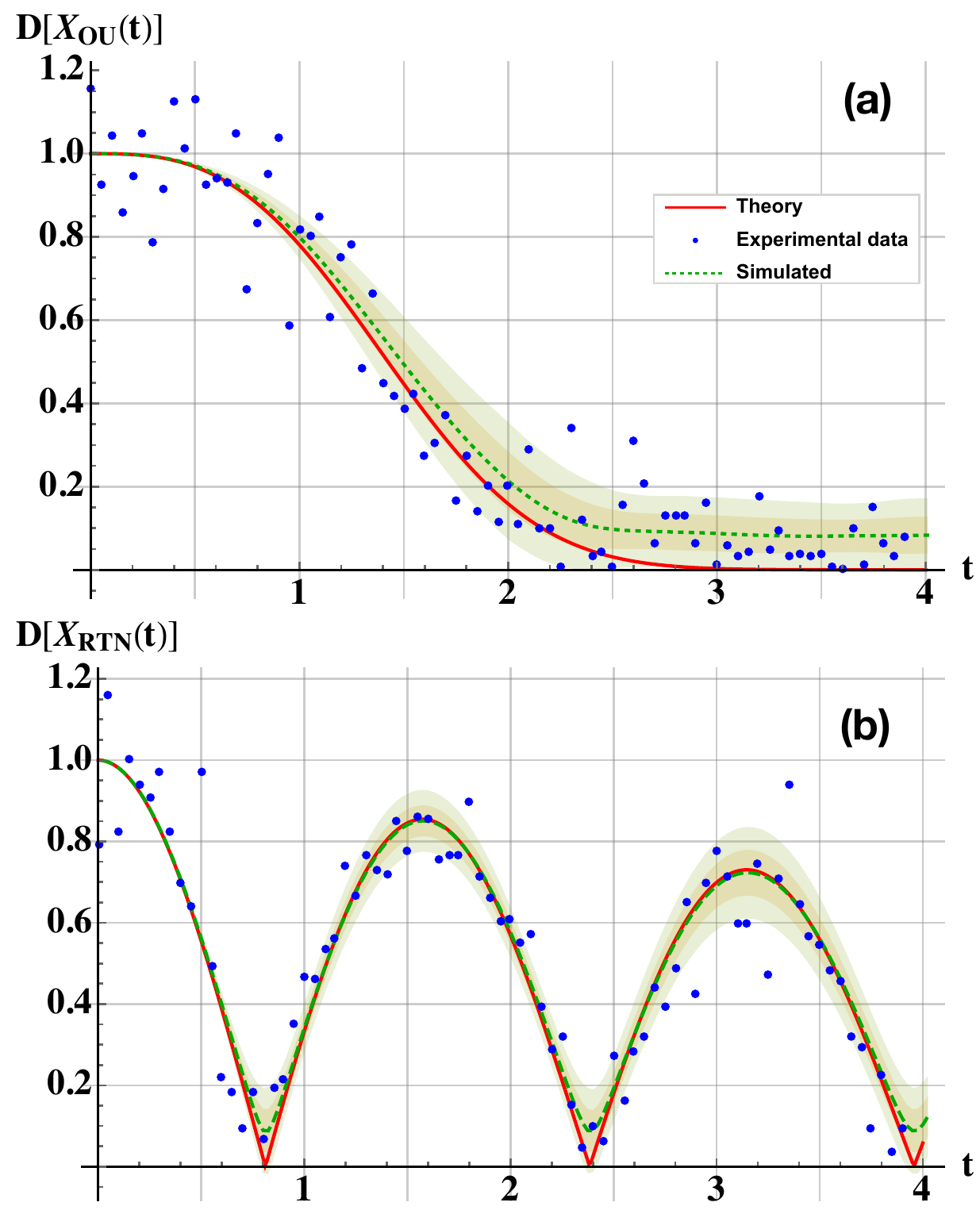}
  \caption{\label{fig:markov}Behavior of the quantum
    non-Markovianity quantifier $D$ defined in
Eq.~(\ref{eq:decoh}) for the
    case of classical Markovian processes. In panel (a) we consider
    the Gaussian Ornstein-Uhlenbeck process $X_{\tmop{OU}}$ with
    $\gamma = 0.1$ and $\sigma = 0.63$;  in panel (b) we consider
    the non-Gaussian but still Markovian random telegraph noise
    process $X_{\tmop{RTN}}$ with $\gamma = 0.1$. 
    Blu dots represent the experimental data and the red line is the analytic solution.
    The green dashed line is the average of 100  simulated curves, each 
    obtained with 100 realizations of the noise. 
   The dashed areas correspond to the  $1\sigma$ (darker) and
   $2\sigma$ (lighter) interval around the averaged coherence and $\sigma$
   is the standard deviation of the sampled curves.
    In the first case (a)
    the quantum dynamics does exhibit a Markovian behavior,
    corresponding to a monotonic decrease of coherence, while for the
    RTN (b) the resulting quantum dynamics is
    non-Markovian.}
\end{figure}
The values obtained in correspondence to
the different realizations have been encoded in different regions of the SLM,
thus allowing for an easy implementation of the average as depicted in the
logical scheme Fig.~\ref{fig:schema}.
Given that the aim of the work is the comparison between
non-Markovianity of the quantum dynamics and the features of the
classical noise, for each kind of noise we have studied the behavior
of the trace distance as a function of time.

\begin{figure}[H]
  \includegraphics[width=0.99\columnwidth]{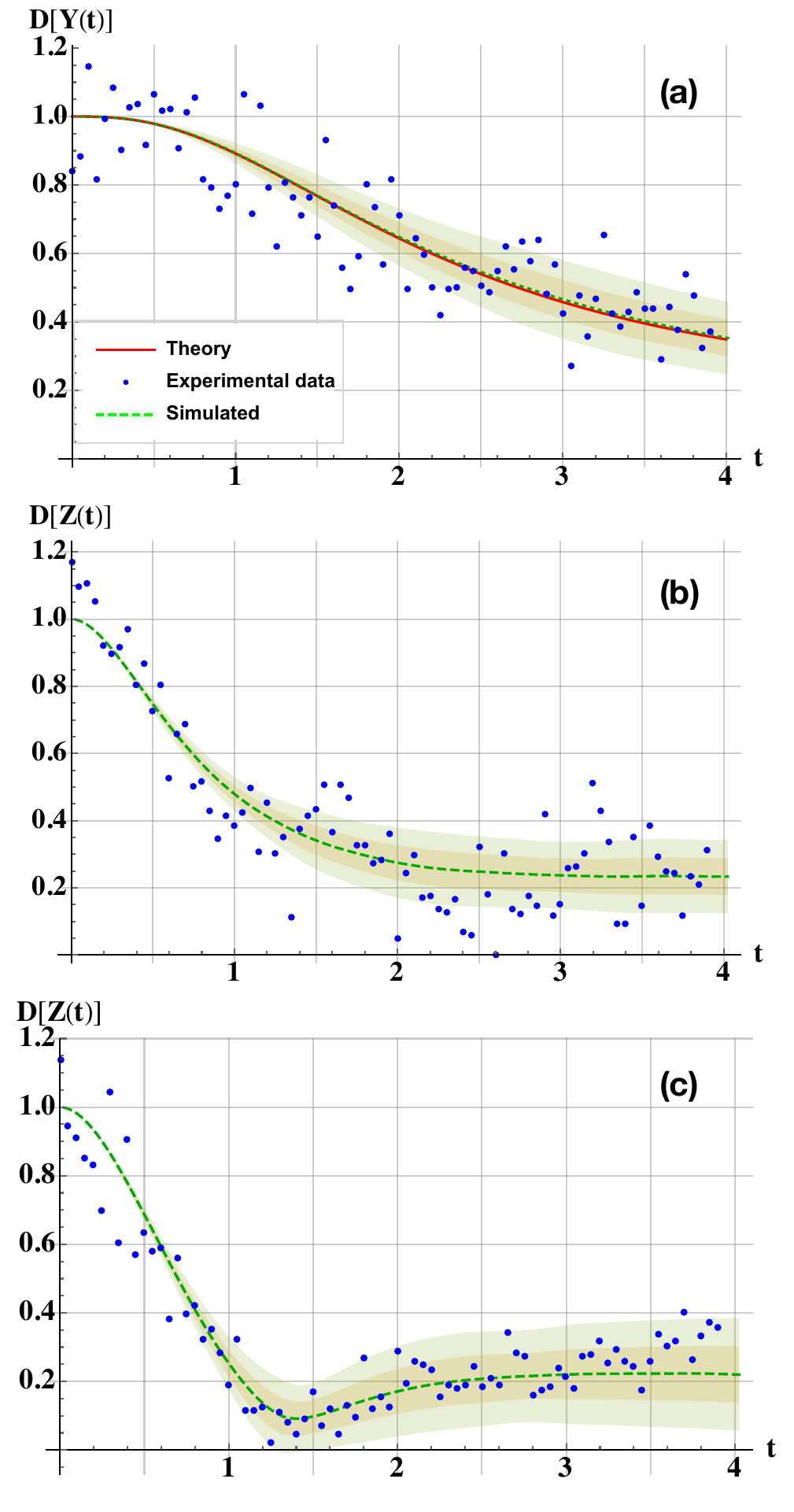}
  \caption{\label{fig:non-markov}
Same  non-Markovianity quantifier $D$ shown in Fig.~\ref{fig:markov}
for the
    case of classical non-Markovian processes. 
    In panel (a) we consider
    the Gaussian but non-Markovian processes $Y(t)$ with $k=1$; in panel (b) we consider
    the non-Gaussian and non-Markovian process $Z(t)$ with $\mu = 1$;
    in panel (c) we consider
    the same process $Z(t)$ with $\mu = 0.5$. 
    In the first two panels
    the quantum dynamics does exhibit a Markovian behavior,
    corresponding to a monotonic decrease of coherence, at variance
    with the classical property. In the last panel, instead, one also
    has quantum revivals corresponding to a non-Markovian behavior.
    As in Fig.~\ref{fig:markov} the blue dots represent the experimental data and
    the red line the analytic solution, when it exists.
    The green dashed line is the averaged non-Markovianity and
    the shaded areas corresponds to $1\sigma$ and $2\sigma$ 
    regions around the mean value.
  }
\end{figure}

We keep track of time by
encoding in the SLM the different values of the processes at
discretized times with step of order $1$ in inverse units of the rate
appearing in the stochastic differential equation characterizing the given process.
As discussed in Sec.~\ref{introTD}, we consider the quantity $D$ defined in
Eq.~(\ref{eq:decoh}) as quantifier of the non-Markovian
features of the dynamics, which in particular fixes the behavior of the
coherences. In Fig.~\ref{fig:markov} we show the experimental data
referring to the quantum signature of non-Markovianity for two
classical Markovian processes, namely Ornstein-Uhlenbeck  $D [ \{
X_{\tmop{OU}} ( t ) \} ]$ and random telegraph noise $D [ \{
X_{\tmop{RTN}} ( t ) \} ]$. While the former quantity is monotonically
decreasing, the latter clearly shows a damped oscillating behavior,
corresponding to a quantum non-Markovian behavior. Note that both
processes have a power spectrum of the form Eq.~(\ref{power}). While both processes are classically
Markovian, only  Ornstein-Uhlenbeck is Gaussian. The theoretical and numerical previsions are in very good agreement with the experimental data (see also the shaded regions in Fig.~\ref{fig:non-markov})

We further consider two non-standard classical processes obtained as
solution of the stochastic differential equations \eqref{eq:ii} and
\eqref{eq:iii} respectively. The process $Y ( t )$ is Gaussian but
classically non-Markovian. Despite these properties and the non
trivial power spectrum given by \eqref{power}, as shown in
Fig.~\ref{fig:non-markov} the quantity $D [ \{ Y( t ) \} ]$ is
monotonically decreasing in time. Again the experimental points are in
agreement with the analytical estimate \eqref{eq:dy}.
In the case of $D [ \{ Z( t )
\} ]$, there exist values for the parameters
that make revivals of the trace distance appear. 
We highlight again that the structured spectrum of 
both the $Y(t)$ and $Z(t)$ processes cannot be directly 
connected to memory effects.
 For such non-Gaussian process the experimental points are
compared to the results obtained via a numerical simulation of the
process, further allowing to obtain its power spectrum shown in
Fig.~\ref{fig:spettroZ}. Again the classical non-Markovianity of the
process is not reflected in the quantum signature.

\section{Conclusions and outlook}
\label{ceo}
We address the quantum non-Markovianity of a single-qubit dephasing
map in terms of the Markovianity of the stochastic process generating
the noise. In particular, we considered four random processes with
different Gaussianity and Markovianity traits.  We showed that the
Markovianity of the classical stochastic process does not affect the
information backflow to the system, i.e. classical lack of Markovianity
is not directly related to memory effects.  However, we showed
evidence that the non-Gaussianity of the noise can be related with
oscillations of the trace-distance.

\section*{Acknowledgements}
The author acknowledges support from the Joint Project ``Quantum Information
Processing in Non-Markovian Quantum Complex Systems'' funded by FRIAS,
University of Freiburg and IAR, Nagoya University, from the FFABR project of
MIUR and from the Unimi Transition Grant H2020. Bassano Vacchini gratefully
acknowledges useful discussions with Alberto Barchielli and Matteo Gregoratti.

\end{document}